# Global Inequalities in the Production of Artificial Intelligence

## A Four-Country Study on Data Work

Antonio A. Casilli[1], DiPLab, Institut Polytechnique de Paris, 91120 Palaiseau, France
Paola Tubaro, CREST, CNRS, Institut Polytechnique de Paris, 91120 Palaiseau, France
Maxime Cornet, i3, Télécom Paris, Institut Polytechnique de Paris, 91120 Palaiseau, France
Clément Le Ludec, CERSA, Université Paris-Panthéon-Assas, 75005 Paris, France
Juana Torres-Cierpe, LaborIA, Inria, 78150 Le Chesnay-Rocquencourt, France
Matheus Viana Braz[2], Maringá State University (UEM), Paraná, Brazil

## INTRODUCTION

The recent emergence of artificial intelligence (AI) as a transformative force across industries has directed attention toward potential job losses, as machines may assume tasks currently performed by workers. Ironically, though, human labor plays an essential albeit largely unrecognized role in the production of AI. In the present contribution, we shift the lens to this often-overlooked facet of technological development and unveil not only its labor-intensive nature, but also its intricate connections with global economic dependencies, ongoing transformations of the forms and conditions of work, as well as digital, social, and economic inequalities.

The human inputs to the production of voice recognition, text generation, computer vision, and other state-of-the art technologies are not limited to the highly paid data scientists and engineers foregrounded in Silicon Valley narratives. The bulk of the workforce underpinning today's AI are data workers (also known as “micro-workers,” “crowdworkers,” and “cloudworkers”) who operate in the shadows. They prepare data to train machine learning algorithms: for example, they annotate traffic images that serve as examples for self-driving cars, indicating what each element represents (say, a pedestrian, a bike, and so on). Other workers will check outputs, to ensure that once in use, a given AI technology functions properly. Workers are also sometimes asked to replace algorithms for tasks that are hard to automate—for example, when subjective appreciation is needed to distinguish offensive from acceptable social media contents (Tubaro et al. 2020a).

Many of these tasks are small and repetitive and do not require the competencies customarily associated with computer engineering or data science. They are therefore considered peripheral to the core value-adding activities of AI producers, and commonly outsourced via online labor platforms and networks of subcontractors. Regardless of the qualifications of data workers and of the actual skills they display in their activity, their contribution is little appreciated, remains invisible to the public and sometimes even to industry actors, and attracts low compensation. With few exceptions, data work is extremely fragmented and payment by piecework is customary.

By examining the roles on these workers and their contribution, we dispel the notion that AI represents a complete departure from labor-intensive processes. But the tech industry does make data work invisible, thereby strengthening a broader ongoing tendency to shift power and resources away from labor and toward capital. Through a comparative study of data workers across four countries, we highlight the complexities of the global transformations of digital labor that influence the production of AI and are in turn reinforced by it. We show how diverse socioeconomic settings, in higher- and lower-income parts of the world, differentially affect AI production and the organization of data work. This comprehensive overview underscores the global dependencies and the digital, economic, and gender-based inequalities that underpin AI in its current form.

## INEQUALITIES IN THE GLOBAL DATA WORK MARKET

Spurred by the launch of Amazon's pioneering Mechanical Turk platform, data work took root in the second half of the 2000s and has been growing steadily since then. Like other forms of digital labor, it relies mainly on platform intermediaries that facilitate connections between businesses on the one side and a diverse pool of flexible labor providers, construed as independent workers, on the other. But unlike other instances of digital labor, where tasks like delivery and personal services are performed offline although intermediation occurs online, data work can be largely done remotely, enabling cross-border matching of clients and workers. Thus, it is often regarded as part of online labor, which also includes qualified freelancing activities such as design, computer programming, and management consulting. Together, online data work and freelancing encompass a whopping 163 million people globally (Kässi et al. 2021). Differences between the two hinge on complexity and scale of tasks, mode of negotiation, and levels of remuneration, all more disadvantageous in data work (Kuek et al. 2015). In particular, the average hourly earnings of freelance platform workers are estimated at US$7.6, and those of data workers at US$3.3 (ILO 2021).

The early literature on data work predominantly reflected the experience of Amazon Mechanical Turk, with its workforce mostly resident in the United States and secondarily in India (75% and 16%, respectively, according to Difallah et al. 2018). Ross et al. (2010) found that more than two-thirds of these workers were under 35 years old and concurrently engaged in other occupations, so that they used data tasks as supplemental sources of income. Women were slightly more numerous than men in the United States, but the ratio was reversed in India, where platforms are more often the main source of income.

In subsequent years, technological advances requiring larger-sized datasets and higher-quality annotations pushed up demand for data work, while the number of platform intermediaries tripled (ILO 2021), resulting in Amazon Mechanical Turk losing relevance compared to competitors. Labor supply expanded even more, with growing participation from a diverse workforce distributed across the globe. A landmark ILO report over five platforms documented data work in 75 countries (Berg et al. 2018). Because most data tasks can be performed from home, the outbreak of the COVID-19 pandemic in 2020 triggered a significant acceleration in its spread worldwide, with platforms like Clickworker surpassing the 2 million registered workers milestone (further increased to 4.5 million at the time of writing this chapter). Massive oversupply of labor, including from low-wage countries, and tight international competition put downward pressure on remunerations. The platform model offers no protection against the fluctuations of the planetary market, especially as clients and intermediaries insist on the unskilled—thus, undeserving—nature of tasks.

There is growing evidence that some data workers are recruited outside of freely accessible online markets. Schmidt (2022) notes the development of a “specialized, full-service crowd-AI stack” after the initial rise of “general-purpose” platforms à la Mechanical Turk, while Miceli et al. (2020) and Miceli and Posada (2022) report cases of long-term contractors that negotiate large deals with clients and hire teams of workers to execute them. Platforms themselves are sometimes managed by one or more vendors on behalf of a single, major technology producer (Gray & Suri 2019). This diversity of business models suggests that the role of platforms and the modes of their organization are more varied than previously believed, and that the size of the data work market surpasses any counts of registered users of the most popular among them. Tracking data work is becoming more challenging.

The contribution of data work to AI has been revealed by Crawford (2021), Gray and Suri (2019), Miceli and Posada (2022), Schmidt (2019), Casilli (2025) and others, highlighting dynamics that both challenge and parallel dire predictions about the future of work. Human labor is important to the production of AI, but it is less visible, less paid, and less protected than conventional employment. Other authors have discussed lack of social protection, hardships induced by algorithmic management, and surveillance during work hours (Aloisi & De Stefano 2022). Issues of worker rights, job security, and equitable compensation persist even in countries with established labor legislation, as AI projects can entail long hours and grueling workloads.

Contrasting advanced industrial economies with lower-income countries, characterized by different labor regulations and technological landscapes, presents a more variegated picture (Surie & Huws 2023). Here, online data work often exists within informal labor (Rani & Furrer 2019). The tasks performed are usually more repetitive and routine. For example, Lindquist (2022) and Grohmann and Fernandes Araújo (2021) show that in Indonesia and Brazil, respectively, data work often occurs in “follower factories” and “click-farms,” where tasks range from creating fake content to “liking” and “sharing” clients' online profiles. Though apparently disruptive of the normal functioning of social media websites, click-farming ultimately contributes to automating search and recommendation algorithms. Just as legally dubious, “captcha sweatshops” serve the purposes of hackers while also supplying labor to train computer vision solutions through the tagging of images (Pettis 2023). In comparison with their counterparts who filter spam, illegal content, and fabricated data, typically gathered under the banner of Commercial Content Moderation, the status of captcha solvers and click farmers is even less prestigious, less recognized, and consequently less paid. According to Roberts (2019), moderators' activity sits on a continuum between tasks performed on data work platforms and “boutique companies” in low- and middle-income countries, and tasks performed in industrial facilities for major technology companies in richer regions.

This complex landscape can be interpreted in light of the cross-country gaps and interdependencies that reflect a larger international division of digital labor (Fuchs 2016). Inherited economic disparities shape the flows of data work, whereby AI producers in industrial countries outsource tasks to providers worldwide, with less-paid and less-prestigious activities being mostly undertaken in lower-income regions. These dynamics intersect with in-country inequalities, both “legacy” gaps rooted in historical socioeconomic structures, and “emergent” ones resulting from the changing nature of work in the digital age (Robinson et al. 2020a, 2020b). Although various dimensions of inequality may matter, space limitations suggest focusing here on just three: first, economic disparities based on income and wealth; second, the so-called digital divide, which we take for simplicity to be the combination of access to, and literacy in, digital technologies, affecting people's ability to perform data work; third, gender differences in terms of access to the labor market and household responsibilities. There is indeed scattered, but growing evidence of gender differences in digital labor just as in other economic activities (Fuster Morell 2022). We posit that workers' positions at the intersection of these three dimensions coevolve with the global data work landscape, and gaps may be exacerbated by its organization as low-income countries engage in lesser-paid routine tasks while high-skilled, higher-paid activities remain concentrated in the minority world. As a result, the global supply chains of AI may increasingly resemble the “poverty chains” that Selwyn (2019) observed in more traditional industries.

## A FOUR-COUNTRY STUDY

The interplay between cross-country economic inequalities, disparities in digital literacy and access, gender differences, and fast-paced technological development, calls for a comprehensive understanding of the implications of AI production. We now present four country studies to illuminate how these inequalities manifest in specific settings, underscoring the far-reaching impact of data work on the broader socioeconomic backdrop. Lower-income regions often find themselves locked into roles of suppliers of cheap labor in the AI value chain, while richer regions occupy the higher tiers. Although the latter too host data workers, they wield significant decision-making power and benefit from the development and utilization of these technologies.

To ensure sufficient variation in the data, our choice of studies starts from apparently peripheral countries that provide data workforce reservoirs but do not host core AI production activities, and then move to a more central one that has both. We foreground Spanish-, Portuguese-, and French-speaking countries, a departure from previous studies that dealt almost always with English-language settings. We also move away from the common practice of studying Amazon Mechanical Turk, which as noted above may not reflect the latest trends in the industry. We leverage original quantitative and qualitative data that we collected between 2018 and 2023 among data workers and other stakeholders, recruited through a sample of platforms and other intermediaries.

We start from Venezuela, whose relevance in the planetary market for data work has already been noted (Johnston 2022; Posada 2022; Schmidt 2022). The severe economic and political crisis that has plagued the country in the last few years has pushed large segments of its highly educated population to international data work platforms, which pay in hard currency and thus—paradoxically—provide higher and more stable income flows than local jobs. We then move to Brazil, a vibrant emerging economy characterized by sharp internal income inequalities and a tradition of informal labor. The size of its data-working population is increasing rapidly and diversifying from click-farming to a range of AI-supporting tasks. We study its various forms and how they meet the needs of disadvantaged groups within the country. Looking at Africa, Madagascar has experience exporting computing services and builds on a history of commercial exchanges with its former colonial ruler, France; however, a steep digital divide requires adaptations for data work to thrive. We discuss the organizational arrangements that have brought data work to the country, and their societal impacts. Finally, we take France as example of an economy that produces advanced AI solutions while also hosting data workers, although their conditions differ from those of their lower-income counterparts.

## Venezuela

Venezuela, a major oil exporter since the 1970s, achieved significant social goals under the presidency of Hugo Chávez in 1999–2013, among them reducing child mortality, extending life expectancy, and expanding education access. Campaigns to distribute computers to young people and to spread digital literacy were successful. But fluctuations in oil prices since the mid-2010s, combined with political instability and sanctions, have plunged the country into a deep economic crisis. Venezuela is still heavily dependent on oil (58% of its budget in 2024) to the detriment of other sectors, and hyperinflation was at 360% in 2023. The US embargo and attempts to destabilize Venezuela have had a huge impact on the country. This has created the conditions for foreign platforms, including the one we looked at in our inquiry, to target Venezuela as a sizable pool of cheap workers. While Venezuela has seen recent modest economic gains, significant challenges remain, with 80% of the population still living in poverty. This is further compounded by high emigration, declining birth rates, and increasing mortality rates (ENCOVI 2022).

In these conditions, data work through international platforms appears as a way forward. Even low-paying tasks like captcha-solving are priced in hard currency and therefore, compare favorably to plummeting local wages. Thus, international platforms have seen massive inflows of Venezuelans since about 2017. To understand their motivations and practices, we launched an online survey (in Spanish) through the platforms Microworkers (2020–2021) and Clickworker (2022), collecting 283 questionnaires and 19 in-depth interviews. This was part of a broader study encompassing the whole of Spanish-speaking Latin America together with Spain, which gathered over 2,400 questionnaires and almost 60 interviews overall.

Two thirds of Venezuelan workers are men, 70% are under 35 years of age, and over half have a university degree, often in disciplines like engineering and computing (Tubaro 2022). For three quarters of them, platform data work is the main source of income, and earnings are primarily used to purchase necessities. They often work full time, starting very early in the morning to synchronize with the business hours of clients (based primarily in the United States, secondarily in Europe). While many of them understand how data work serves AI development, they must surmount major obstacles to undertake it. Aging computing equipment, slow internet connections, and frequent power outages limit the types of tasks they can do, and increase the risk of mistakes—thus, the risk of not getting paid. Financial restrictions require use of complex strategies—from recourse to the black market to investment in cryptocurrencies—to convert online earnings into local currency. Finally, despite their willingness to accept even very poorly paid tasks, Venezuelans face tight competition both from their numerous compatriots and from other low-income countries.

Nevertheless, workers actively cooperate to share the benefits of platform work, particularly through online groups that help members identify suitable tasks, understand clients' demands, and even just socialize with one another (see also similar patterns in Soriano's chapter in this volume on the Philippines). The 18-year-old male founder of one such group commented that he aimed “to assist other people who were joining the Microworkers platform, both to offer help and receive it, by providing data, information, and opinions regarding platforms that can help generate more income.” To an extent, workers take on mutual support and personal/professional development roles that in conventional workplaces are the employer's purview.

## Brazil

Brazil is the seventh most populous country in the world, characterized by sharp income inequalities, high unemployment rates especially among young people, and a large informal economy that employs almost 40% of the active workforce. In recent years, the country has been undergoing progressive erosion of labor and social rights, parallel to the growth of digital platforms. Both trends shift risks and costs onto workers and contribute to the long-term process of informalization of work. More than 50 international platforms are used by Brazilian data workers (Viana Braz 2021) to contribute to AI development. To shed light on the profiles and activities of these workers, we conducted a digital ethnography across more than 20 online groups focused on data work, and in-depth interviews with 15 workers active on various platforms (2022). We then collected 477 questionnaires from Brazilian users of the platform Microworkers (2023). The questionnaire was adapted from the Venezuelan one and translated into Portuguese.

Brazilian data workers are as young as their Venezuelan counterparts, but unlike the latter, almost two-thirds are women, and only about one-third relies solely on data work platforms as their source of income. As many as 40% are unemployed, without a professional activity, or in informal work. Indeed, informality is a distinctive feature of these workers' trajectories, as half of them have experienced it at least twice in their careers. Even those in formal work are often part-time or have changed jobs frequently. On average, family income is about 30% lower than the general population of the country. Women are more numerous among the unemployed, and more often have childcare responsibilities. Educational attainments are more diverse than in Venezuela, those with university degrees being just as numerous as those with secondary school only. These differences do not prevent the emergence of solidarity among Brazilian data workers, although only about one in five uses online groups. Mutual help can also occur offline: for example, an interviewee recalled having sought help from a neighbor to take pictures that she had to upload to her data work platform.

A first cross-country comparison exposes two different pictures. If in Venezuela, data work is a collective undertaking that involves large segments of the population, all equally affected by the crisis, it is driven by young men with science and technology backgrounds, who invest a lot of their time and know-how to reap the largest benefits from what has become their (and their acquaintances') main earning activity. In richer, but more unequal Brazil, data work provides a complementary source of income to a subgroup of digitally literate and equipped people, who are nevertheless (partly) left out of formal labor markets and must rely on informality to make ends meet. Women with children, people with nonlinear career trajectories, as well as low-wage and less-educated workers are overrepresented within this group, and thus make the bulk of the data workforce for AI. Either way, technology producers from North America and Europe harness the relative deprivation of specific populations—a whole country in one case, and a disadvantaged subgroup in the other—to access low-cost data work for their purposes.

## Madagascar

Though endowed with considerable natural resources, Madagascar has one of the world's highest poverty rates owing to weak infrastructure, repeated natural disasters, and limited access to health and education services. Nevertheless, it is the second largest exporter of computing services in Francophone Africa, despite the widespread informality and small size of the sector, which accounts for only 3% of GDP. Its dynamism is the result of policies that, since the 1990s, have encouraged export-oriented activities. In this context, data work has arrived as an addition to the set of services that local businesses offer to their foreign (French, and to a lesser extent European and Asian) clients and that encompass, among others, telephone customer support, remote sales, and data entry.

Our 2021–2022 fieldwork in Madagascar revealed the continuity between today's data work and older call centers and teleservices, in particular how AI production relies on the same offshoring dynamics that structure the value chains of more mature industries. It also uncovered a diverse array of organizational structures that engage in data work. Here, gaps in internet access prevent work-from-home for many people, thereby curbing the diffusion of international platforms like Microworkers and Clickworker, widely used in Venezuela and Brazil. In Madagascar, we mostly investigated small local companies, both formal and semi-informal, which hire workers to execute (usually large) contracts for French technology startups (Le Ludec et al. 2023). Overall, we interviewed 147 workers and managers of 10 companies offering data work services, and we administered 296 questionnaires to workers. For comparison purposes, we also did a few interviews with workers in Cameroon, Côte d'Ivoire, and Senegal, and with company managers in Egypt.

Nine out of ten employees of these companies are under 35 years of age, three quarters have higher education degrees, and over two-thirds are men, who often see data work as their first step into the formal economy after university. Though most workers enjoy better social protection and wages than in the informal sector, around half of their pay is a bonus tied to performance metrics—a hybrid payment structure that introduces income volatility and echoes historical manufacturing practices. Interviewees complain that their remunerations are low compared to the cost of living in the capital, especially for young families with children; that it is difficult to save for the future; and that wide pay gaps persist between formal and informal workers. Interestingly, the hierarchical organization of companies does not facilitate career progressions, although horizontal transitions (say, from annotation to content moderation) are common. Indeed, French clients maintain tight control over their production processes and leave very few intermediate management positions available to locals. Unsurprisingly, workers perceive their current roles as temporary and hope to transition to jobs more aligned with their training—although the local labor market stagnates, and limited capital availability thwarts many entrepreneurial ambitions.

The ultimate beneficiaries are their overseas clients, which can tap into a skilled, digitally savvy workforce at lower cost. By exerting control over their contractors' work processes, AI companies effectively manage a "quasi-employee" relationship without the obligations associated with formal employment under French law.

## France

If France is the destination of important flows of data work from Madagascar and other parts of Africa, it is itself a reservoir of data workers, and it was the country where we started our investigations back in 2018. An advanced industrial country with over 90% internet penetration and very high equipment rates, France invests heavily in AI development, and its firms are more numerous to adopt data work services than their German counterparts (Belletti et al. 2021). France has also seen a proliferation of data work platforms (23 in 2019, over half of them local), with participation of around 260,000 workers (Tubaro et al. 2020b) who do tasks that require local language and culture, and therefore cannot be easily offshored. Here, platforms advertise themselves as leisure-adjacent, as sources of extra income, and as occupations at the forefront of technology. To investigate them, we collected data from a Paris-based platform that only recruits French residents, then called FouleFactory and later renamed Yappers, with 909 online questionnaires and 90 interviews, mostly with workers and secondarily with platform managers, clients, and other stakeholders.

French data workers are mostly women, often with kids, confirming a trend observed in Brazil, but not in Venezuela and Madagascar. Workers are older than in the three other countries, with almost two-thirds between 25 and 44. Qualifications are as high as elsewhere, with two out of five workers having university degrees. Over two-thirds use data work as a supplementary source of income, usually in addition to a main salaried job (more often part-time for women). But French data workers are overrepresented among those in inactivity and unemployment, and more than one in five live below the poverty line. Platforms may claim that data work is a fun hobby for techies, but they cater to disadvantaged groups, particularly women who are digitally savvy, old enough to have a family, but not earning enough to support it. Precarity persists along lines of gender inequality inherited from the conventional economy, as women need to do more data work (because their part-time jobs pay too little) but have less time to search and select the best tasks online (because they have more childcare responsibilities). Opportunities to transition to better-paying occupations in the technology sector are also more limited for women, fewer of whom hold degrees in science and technology and who have less social capital in terms of contacts to professionals who could advise them in these areas (Tubaro et al. 2022). Data workers in this country experience the highest levels of isolation and are the least likely to use online groups for active solidarity.

## TRANSNATIONAL DYNAMICS IN DATA WORK AND THE PERSISTENCE OF HISTORICAL DISPARITIES

Inequalities manifest in multifaceted ways across low- and high-income countries, mainly influenced by gender and digital divide, in ways that take on distinct contours in each setting. Venezuela and Madagascar exemplify the reality of low-income countries, where data work for foreign clients is popular among large segments of the population. Digitally knowledgeable and well-educated young men are at the forefront, practice data work as a full-time activity, and derive their main income from it. Venezuela accesses the international online market for data tasks directly via international platforms, but it is limited by gaps in access to equipment and internet connections, which, together with strong competitive pressure, keep remunerations down. Madagascar bypasses its internet infrastructure problems through small companies that provide equipment and connection to workers, while also mitigating income volatility through (more or less formal) employment contracts; however, its asymmetric relationships with clients prevent it from moving up the AI supply chain.

In richer Brazil and even more in France, data tasks are residual activities more commonly undertaken by disadvantaged niches within the population, especially low-earning workers and women with children, who need a complementary source of income. For Brazil's young, jobless new graduates, data work offers an alternative to traditional informal jobs, although its role as a springboard to better future jobs remains to be proven. Opportunities for professional development may seem stronger in France owing to its well-functioning formal labor market, widespread internet access, and burgeoning tech sector. However, gender disparities, correlated with gaps in access to advanced digital literacy and social capital, hinder women's progression prospects.

All four cases show that AI producers tap into pools of disadvantage to find providers of affordable data work. Although the sources of disadvantage vary within each country, data workers represent a specific (more or less large) group at the intersection of economic, digital, and gender-based gaps, both across and within nations. The skills and contributions of these workers are barely recognized, as the structure and international extension of the AI supply chains, through platforms and other intermediaries, often across borders, keep remunerations down and fail to leave space for career progression opportunities. There are also differences across the four countries, depending on their income levels. AI-driven activities are found in high-, middle-, and low-income countries, where a common thread of poorly paid and precarious data work is evident. However, the proportion of skilled, highly paid technology workers to data workers serves as an indicator of a country's position within the supply chain. All other things being equal, a higher ratio means a more advantageous position, as for example in France.

To generalize from these cases and see the overarching connections between them, we now merge our results with evidence from existing literature to schematically represent the network that interconnects data work providers and AI-producing clients on a global scale. The visualization presented in Figure 12.1 aims to illustrate the flow of data work across regions, unveiling patterns of concentration, directionality, and historical influence.

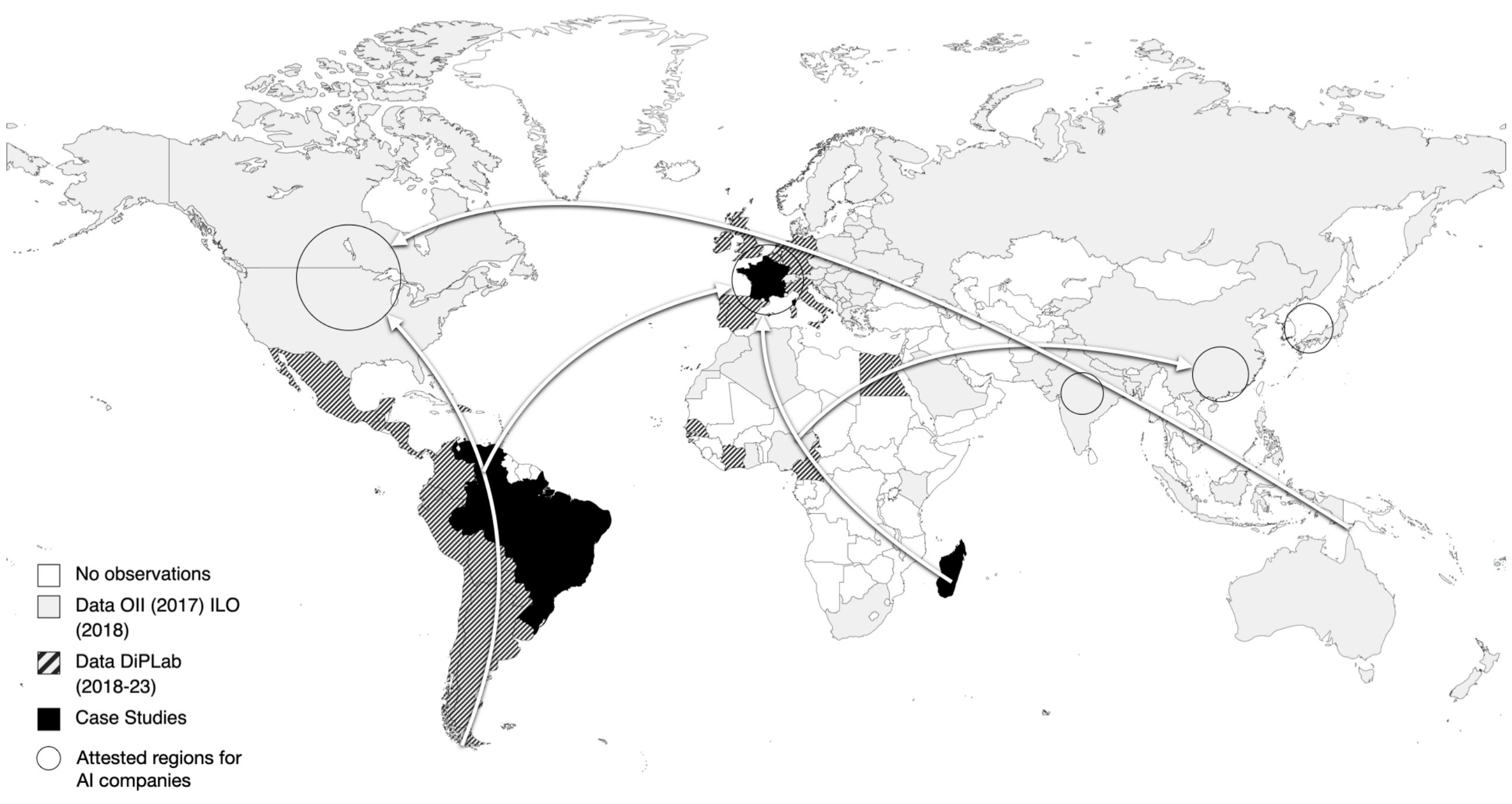

Figure 12.1 Global flows of data work for AI production, showcasing three main directions and their interconnections between data work providers and AI producers in different parts of the world. The East-to-West flow documented in the literature originates in South-East Asia and progresses to Europe, the United Kingdom, and most often the United States. The South-to-North flows involve routes from Africa to Europe, with a spinoff to China via the Gulf countries, and from Latin America and Mexico to North America, with an accompanying European spinoff. Source: authors' elaboration

AI producers (i.e., companies that market AI solutions while acting as clients for data work providers) are concentrated in North America and Europe, with growing presence of China and India. On the labor supply side, data work predominantly emanates from the Global South. The diagram uncovers three main flows:

East-to-West. Emerging from South and South-East Asia, particularly India, Bangladesh, Pakistan, and the Philippines, it traverses westward. Encompassing Europe and the United Kingdom, it culminates in the United States, which additionally boasts a large internal market for data work.

South-to-North (with West-East spinoffs). One flow connects Africa to Europe. It originates in Madagascar and other countries such as Côte d'Ivoire, Cameroon, and Senegal. This flow reaches European AI producers, passing through Morocco and Tunisia where many intermediaries (in charge of administrative, technical, and management tasks for European clients) are located. A spinoff route connects parts of Africa to the Gulf countries and China. The other flow links Latin America

> to North America. It begins in Venezuela, Brazil, and other countries such as Argentina, Colombia, and Mexico, and eventually reaches the United States. An ancillary trajectory channels data work to Europe.

This global map highlights the need to shift our gaze toward the Global South, where data work takes diverse forms shaped by historical legacies, socioeconomic disparities, and geopolitical positions. Countries such as Venezuela, Brazil, and Madagascar are hubs for labor-intensive activities, drawing on educated but precarious youth who engage with international labor markets through AI-related data tasks. The Global South's contributions to AI production transcend mere cost-effectiveness, unveiling intricate organizational models that challenge simplistic narratives. This map also reveals that in contemporary globalized AI production, vestiges of colonial and postcolonial relationships continue to influence the present. In the mid-18th century, the British East India Company took control, and later direct rule, over India. The Philippines holds historical colonial links with the United States, from the time of colonization in the first half of the 20th century. Several African countries share a past of colonial domination by European powers, which have contributed to lasting cultural and socioeconomic ties. Latin American countries' ties to Europe, primarily Spain and Portugal, stem from centuries of dependency, shaping their languages, cultures, and socioeconomic systems.

Despite the technological veneer, established power dynamics persist in the digital era. The practice of offshoring labor-intensive tasks to the Global South reverberates with echoes of past economic exploitation, thereby reinforcing existing inequalities and entrenching power dynamics. The asymmetrical geography that emerges, mirroring historical patterns, is examined in the digital labor literature under the conceptual framework of “data colonialism” (Couldry & Mejias 2019) and “digital coloniality” (Casilli 2017). Nevertheless, the dependencies and imbalances in power and wealth that permeate AI production and its supply chains are not straightforward reproductions of historical forms of imperialistic domination and economic extraction. While certain instances reveal one-to-one replication of age-old colonial models, such as the linkages of French-speaking African countries with French AI producers, more complex geopolitical processes emerge. Notably, what can be described as a “Digital Monroe Doctrine” consolidates Latin American countries' dependence on North America, with an influence of political factors in shaping these connections. Several South American countries, such as Venezuela discussed earlier, are affected by US foreign policy that has destabilized and marginalized their economies. Recent linkages of African countries to China further diversify the landscape. Historical legacies and contemporary influences are thus equally important.

Platforms and data outsourcing operations are heavily influenced by the policies implemented by the countries in which they are active. Across the four nations examined in our study, approaches differ greatly spanning from laissez-faire to economic incentives. Due to embargoes and animosity toward North American companies, Venezuela doesn't offer specific incentives for platforms. In terms of regulation, they're no more important than any other unlicensed means of obtaining income that its impoverished population might devise. Madagascar, on the other hand, sees data work as information process outsourcing and actively seeks investment. The country has implemented special tax exemption zones under the Free Zone regime, which offer fiscal advantages. The goal is to help boost specific industries, like telecommunications and telemarketing, as well as data annotation and AI training. In contrast, Brazil and France consider these platforms less relevant to their economic growth and regulate them minimally, typically focusing on tax compliance.

## CONCLUSION

A pivotal result stemming from this four-country study is the significant role played by non-English speaking countries in shaping crucial facets of global AI production. Though largely neglected in past scholarship, the cases at hand illuminate the contributions, challenges, and experiences of countries outside anglophone influence. Venezuela, Brazil, Madagascar, and other apparently peripheral countries do shape the trajectories of even US-based AI. By examining data work in four countries, this study has illustrated the interplay of historical legacies and technological advances in the worldwide AI supply chain. Our understanding of the global hold of AI must be recast, highlighting the symbiotic relationship between human intelligence embodied in digital labor practices and machine learning algorithms, thereby contributing to countering the notion of AI as an entirely autonomous entity. The future of work in the digital age requires a broader view, one that accounts not just for the impact of automation on labor, but also for the effects of poor working conditions, coloniality, and inequalities on AI development.

## REFERENCES

Aloisi, A. and De Stefano, V. (2022). Your Boss Is an Algorithm: Artificial Intelligence, Platform Work and Labor. London: Bloomsbury Publishing.

Belletti, C., Erdsiek, D., Laitenberger, U., et al. (2021). Crowdworking in France and Germany (Report 21-09). Mannheim: ZEW.

Berg, J., Furrer, M., Harmon, E. et al. (2018). Digital Labor Platforms and the Future of Work: Towards Decent Work in the Online World [Report]. Geneva: ILO.

Casilli, A. A. (2025). Waiting for Robots: The Hired Hands of Automation. Chicago, IL: University of Chicago Press.

Casilli, A.A. (2017). Digital labor studies go global: Toward a digital decolonial turn. International Journal of Communication 11: 3934–3954. https://ijoc.org/index.php/ijoc/article/view/6349

Couldry, N. and Mejias, U.A. (2019). The Costs of Connection: How Data Is Colonizing Human Life and Appropriating It for Capitalism. Redwood, CA: Stanford University Press.

Crawford, K. (2021). Atlas of AI: Power, Politics, and the Planetary Costs of Artificial Intelligence. New Haven, CT: Yale University Press.

Difallah, D., Filatova, E., and Ipeirotis, P. (2018). Demographics and dynamics of Mechanical Turk workers. In: Proceedings of the Eleventh ACM International Conference on Web Search and Data Mining, 135–143. New York: Association for Computing Machinery. https://doi.org/10.1145/3159652.3159661.

ENCOVI (Encuesta Nacional de Condiciones de Vida) (2022). Universidad Católica Andrés Bello. https://www.proyectoencovi.com/ (accessed 20 August 2023)

Fuchs, C. (2016). Digital labor and imperialism. Monthly Review 67 (8): 14–24.

Fuster Morell, M. (2022). The gender of the platform economy. Internet Policy Review 11 (1). https://doi.org/10.14763/2022.1.1620.

Gray, M.L., & Suri, S. (2019). Ghost Work: How to Stop Silicon Valley from Building a New Global Underclass. Boston, MA: Houghton Mifflin Harcourt.

Grohmann, R., & Fernandes Araújo, W. (2021). Beyond Mechanical Turk: The work of Brazilians on global AI platforms. In: AI for Everyone? Critical Perspectives (ed. P. Verdegem), 247–266. London: University of Westminster Press.

ILO (International Labor Organization) (2021). World Employment and Social Outlook 2021: The Role of Digital Labor Platforms in Transforming the World of Work. Geneva: ILO.

Johnston, H. (2022). In search of stability at a time of upheaval: Digital freelancing in Venezuela. In: Digital Work in the Planetary Market (ed. M. Graham and F. Ferrari), 157–173. Cambridge, MA: The MIT Press.

Kässi, O., Lehdonvirta, V., and Stephany, F. (2021). How many online workers are there in the world? A data-driven assessment. Open Research Europe 1: 53. https://doi.org/10.12688/openreseurope.13639.4.

Kuek, S.C., Paradi-Guilford, C.M., Fayomi, T., et al. (2015). The Global Opportunity in Online Outsourcing. Washington, DC: World Bank Group.

Le Ludec, C., Cornet, M., and Casilli, A.A. (2023). The problem with annotation: Human labour and outsourcing between France and Madagascar. Big Data & Society 10 (2): 20539517231188723. 10.1177/20539517231188723.

Lindquist, J. (2022). "Follower factories" in Indonesia and beyond: automation and labor in a transnational market. In: Digital Work in the Planetary Market (ed. M. Graham and F. Ferrari), 59–75. Cambridge, MA: The MIT Press.

Miceli, M. and Posada, J. (2022). The data-production dispositif. Proceedings of the ACM on Human-Computer Interaction 6 (CSCW2): 1–37. 10.1145/3555561.

Miceli, M., Schuessler, M., and Yang, T. (2020). Between subjectivity and imposition: Power dynamics in data annotation for computer vision. Proceedings of the ACM on Human-Computer Interaction 4 (CSCW2): 115. 10.1145/3415186.

Pettis, B.T. (2023). reCAPTCHA challenges and the production of the ideal web user. Convergence 29 (4): 886–900 10.1177/13548565221145449.

Posada, J. (2022). Embedded reproduction in platform data work. Information, Communication & Society 25 (6): 816–834 10.1080/1369118X.2022.2049849.

Rani, U., and Furrer, M. (2019). On-demand digital economy: Can experience ensure work and income security for microtask workers? Jahrbücher für Nationalökonomie und Statistik 239 (3): 565–597. 10.1515/jbnst-2018-0019.

Roberts, S.T. (2019). Behind the Screen: Content Moderation in the Shadows of Social Media. New Haven, CT: Yale University Press.

Robinson, L., Schulz, J., Blank, G., et al. (2020a). Digital inequalities 2.0: Legacy inequalities in the information age. First Monday 25 (7): 1–27. 10.5210/fm.v25i7.10842.

Robinson L., Schulz J., Dunn H.S., et al. (2020b). Digital inequalities 3.0: Emergent inequalities in the information age. First Monday 25 (7). 10.5210/fm.v25i7.10844.

Ross, J., Irani, L., Six Silberman, M. et al. (2010). Who are the crowdworkers? Shifting demographics in Mechanical Turk. In: CHI'10 Extended Abstracts on Human Factors in Computing Systems, 2863–2872. https://doi.org/10.1145/1753846.1753873.

Schmidt, F.A. (2019). Crowdproduktion von Trainingsdaten: Zur Rolle von Online-Arbeit beim Trainieren autonomer Fahrzeuge. Düsseldorf: Hans-Böckler-Stiftung.

Schmidt, F.A. (2022). The planetary stacking order of multilayered crowd-AI systems. In: Digital Work in the Planetary Market (ed. M. Graham and F. Ferrari), 137–155. Cambridge, MA: The MIT Press.

Selwyn, B. (2019). Poverty chains and global capitalism. Competition & Change, 23 (1): 71-97. 10.1177/1024529418809067.

Surie, A. and Huws, U. (ed.) (2023). Platformization and Informality: Pathways of Change, Alteration, and Transformation. London: Palgrave Macmillan.

Tubaro, P. (2022). Learners in the loop: Hidden human skills in machine intelligence. Sociologia del lavoro 163: 110-129. 10.3280/SL2022-163006

Tubaro, P., Casilli, A.A., and Coville, M. (2020a). The trainer, the verifier, the imitator: Three ways in which human platform workers support artificial intelligence. Big Data & Society 7 (1): 2053951720919776. 10.1177/2053951720919776.

Tubaro, P., Casilli, A.A., Coville, M., et al. (2022). Hidden inequalities: The gendered labor of women on micro-tasking platforms. Internet Policy Review 11 (1). https://doi.org/ 10.14763/2022.1.1623.

Tubaro, P., Le Ludec, C., and Casilli, A.A. (2020b). Counting 'micro-workers': Societal and methodological challenges around new forms of labour. Work Organisation, Labour & Globalisation 14 (1): 67–82. 10.13169/workorgalaboglob.14.1.0067.

Viana Braz M. (2021). Heteromação e microtrabalho no Brasil. Sociologias 23 (57):134–172. 10.1590/15174522-111017.

---

[1] Corresponding author: antonio.casilli@ip-paris.fr

[2] This chapter is based on empirical data collected as part of the DiPLab (Digital Platform Labor) research program. Data collection was undertaken within the following research projects: OPLa ("Organizing Platform Labor"), with funding by IRES-FO, MSH Paris Saclay, and France Stratégie, 2017–2019; HUSH ("The human supply chain behind smart technologies"), funded by ANR, 2020–2023; TRIA ("Le Travail de l'Intelligence Artificielle") funded by CNRS and MSH Paris-Saclay, 2020–2022. Fieldwork in Brazil has also benefited from funding by Minas Gerais Research Foundation (FAPEMIG), through its FAPEMIG/UEMG program (05/2021).